\documentclass{sig-alternate-05-2015}

\usepackage{subfigure}
\usepackage[american]{babel}
\begin{document}

\setcopyright{rightsretained}



\conferenceinfo{Neu-IR '16 SIGIR Workshop on Neural Information Retrieval}{July 21, 2016, Pisa, Italy}


%
\conferenceinfo{Neu-IR '16 SIGIR Workshop on Neural Information Retrieval}{July 21, 2016, Pisa, Italy}

\title{A Study of MatchPyramid Models on Ad-hoc Retrieval}
\subtitle{}
%
%
%
%
%

\numberofauthors{1} 
%
\author{
%
%
\alignauthor {Liang Pang\hspace*{0.5cm}Yanyan Lan\hspace*{0.5cm}Jiafeng Guo\hspace*{0.5cm}Jun Xu\hspace*{0.5cm}Xueqi Cheng}\\
\affaddr{CAS Key Lab of Network Data Science and Technology,\\
Institute of Computing Technology, Chinese Academy of Sciences} \\
\email{pangliang@software.ict.ac.cn, \{lanyanyan, guojiafeng, junxu, cxq\}@ict.ac.cn }
}
\date{30 July 1999}

\maketitle
\begin{abstract}
Deep neural networks have been successfully applied to many text matching tasks, such as paraphrase identification, question answering, and machine translation. Although ad-hoc retrieval can also be formalized as a text matching task, few deep models have been tested on it. In this paper, we study a state-of-the-art deep matching model, namely MatchPyramid, on the ad-hoc retrieval task. The MatchPyramid model employs a convolutional neural network over the interactions between query and document to produce the matching score. We conducted extensive experiments to study the impact of different pooling sizes, interaction functions and kernel sizes on the retrieval performance. Finally, we show that the MatchPyramid models can significantly outperform several recently introduced deep matching models on the retrieval task, but still cannot compete with the traditional retrieval models, such as BM25 and language models.
\end{abstract}

%
%

%
%

%
%



\section{Introduction}
Many text based applications, such as paraphrase identification, question answering, and ad-hoc retrieval, can be formalized as a matching task~\cite{DeepMatch}. Recently, a variety of deep neural models have been proposed to solve such kind of text matching tasks. However, most proposed deep matching models were only evaluated on the natural language processing tasks such as paraphrase identification and question answering~\cite{uRAE, CNTN}. Few deep model has been tested on the ad-hoc retrieval task. Even the models proposed for the Web search tasks, including DSSM~\cite{DSSM} and CDSSM~\cite{CDSSM0}, were only tested on the <query, doc title> pairs which are not a typical ad-hoc retrieval setting.

In this paper, we propose to study the performance of deep matching models on the ad-hoc retrieval task. We choose a recently introduced deep matching model, namely MatchPyramid \cite{MatchPyramid}, which has been shown state-of-the-art performances on several text matching tasks. In MatchPyramid, local interactions between two texts are first built based on some basic representations (e.g., word embeddings). The local interactions represented by a matching matrix is then viewed as an image, and a convolutional neural network (CNN) is employed to learn hierarchical matching patterns. Finally, the high-level matching patterns are fed into a multi-layer perceptron to produce the matching score between the two texts. The model is shown to be able to capture different levels of text matching patterns, such as n-grams and un-ordered n-terms, to improve the matching performance.

\begin{figure*}
  \includegraphics[width=0.90\textwidth]{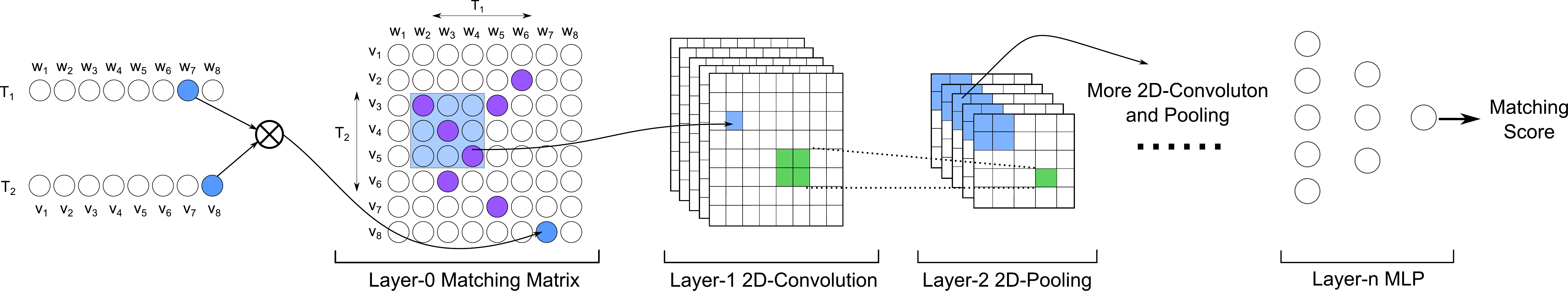}
  \caption{Model structure of MatchPyramid.\label{Figure.model}}
\end{figure*}

When we apply the MatchPyramid model on the ad-hoc retrieval task, we conducted extensive experiments to study the impact of different kernel sizes, pooling sizes and interaction functions on the retrieval performance. We find that three key settings are helpful for the ad-hoc retrieval task, i.e. pooling by paragraph length in document, a good similarity function which can differentiate exact matching signals from semantic matching signals, and a relative small kernel size. Finally, we show that the MatchPyramid models can significantly outperform several recently introduced deep matching models on the retrieval task, but still cannot compete with the traditional retrieval models, e.g. BM25 and language models. The recent proposed deep matching models cannot well fit the ad-hoc retrieval task right now, and more investigation need to be conducted in this direction.

\section{MatchPyramid}
The MatchPyramid model (Figure~\ref{Figure.model}) has three parts, Matching Matrix, Hierarchical Convolution and Matching Score Aggregation.

	\subsection{Matching Matrix}
	In order to keep both the word-level similarity and the matching structures, MatchPyramid introduces the Matching Matrix structure. Matching Matrix is a two-dimension structure where each element $\mathbf{M}_{ij}$ denotes the similarity between the $i$-th word $w_i$ in the first piece of text and the $j$-th word $v_j$ in the second piece of text.
	\begin{equation}
		\mathbf{M}_{ij} = w_i \otimes v_j,
	\end{equation}
		where $\otimes$ stands for a general operation to obtain the similarity.
	
	We define four kinds of similarity functions based on words $w_i$ and $v_j$, or their embeddings $\vec{\alpha_i}$ and $\vec{\beta_j}$.
	
	\textbf{Indicator Function} produces either 1 or 0 to indicate whether two words are identical.
	\begin{equation}
		\mathbf{M}_{ij}=\mathbb{I}_{\{w_i=v_j\}}=\left\{\begin{aligned}
		1, & \qquad\text{if } w_i=v_j\\
		0, & \qquad\text{otherwise.}
		\end{aligned}\right.
	\end{equation}
	
	\textbf{Cosine} views angles between word vectors as the similarity, and it acts as a soft indicator function.
    \begin{equation}
    	\mathbf{M}_{ij}=\frac{\vec{\alpha_i}^\top \vec{\beta_j}}{\|\vec{\alpha_i}\|\cdot\|\vec{\beta_j}\|},
    \end{equation}
    where $\|\cdot\|$ stands for the $\ell_2$ norm of a vector.

    \textbf{Dot Product} further considers the norm of word vectors, as compared to cosine.
    \begin{equation}
        \mathbf{M}_{ij}= \vec{\alpha_i}^\top \vec{\beta_j}.
    \end{equation}

    \textbf{Gaussian Kernel} is a well-known similarity function. This similarity function is introduced in this work based on our studies.
    \begin{equation}
    	\mathbf{M}_{ij}= e^{-\|\vec{\alpha_i} - \vec{\beta_j}\|^2}.
    \end{equation}
    We name MatchPyramid with indicator function as MP-Ind for short. Similarly, we use MP-Cos for cosine, MP-Dot for dot product and MP-Gau for Gaussian kernel.
	
	\subsection{Hierarchical Convolution}
	Based on the Matching Matrix, MatchPyramid conducts hierarchical convolution to extract matching patterns. Hierarchical convolution consists of convolutional layers and dynamic pooling layers, which are commonly used in CNN (such as AlexNet, GoogLeNet) for image recognition tasks.
	
	Kernel sizes in each convolutional layer are the major hyper parameters. In text processing, the size of the kernel determines the number of words we want to compose together. In other words, the kernel size decides the maximum size of n-term features we take into account.
	
	Besides, pooling sizes in each pooling layer are also important hyper parameters, which decide how large the area we want to take as a unit.
	
	\subsection{Matching Score Aggregation}
	After hierarchical convolution, two additional fully connected layers are used to aggregate the information into a single matching score. In this paper, we use 128 hidden nodes for the fully connected hidden layer and ReLU as the activation function.
	
\subsection{Training}
	We use pairwise ranking loss in the training phase. Given a triple $(q, d^+, d^-)$ where the matching score of $(q, d^+)$ should be higher than that of $(q, d^-)$, the loss function is define as:
	\begin{equation}
		L(q, d^+, d^-; \Theta) = \max(0, 1 - S(q, d^+) + S(q, d^-)).
	\end{equation}
where $S(q, d)$ denotes the predicted matching score for $(q, d)$, and $\Theta$ includes the parameters for the feed forward matching network and those for the term gating network. The optimization is relatively straightforward with standard backpropagation. For regularization, we find that the early stopping~\cite{giles2001overfitting} strategy works well for our model.

\section{Experiments}
	\subsection{Dataset and Settings}
	To conduct experiments, we use TREC collection Robust04, which is a news dataset. The topics are collected from TREC Robust Track 2004. The statistics of the data collection are shown in Table~\ref{Table.1}. In this paper, the titles of the TREC topic are treat as the queries. We use the Galago Search Engine in this experiment, and both queries and documents are white-space tokenized, lower-cased, and stemmed during indexing and retrieval.
	
	All the models share the word embeddings trained on the Wikipedia corpus under 50 dimensions. We adopt Adam algorithm~\cite{kingma2014adam} for model training. The learning rate is set to $10^{-4}$. All the MatchPyramid models have one convolutional layer and one dynamic pooling layer, since more convolutional layers and pooling layers will lead to overfitting due to the limited training data.
	
	\begin{table}[!htbp]
		\centering
		\caption{Statistics of the TREC collection Robust04.\label{Table.1}}
		\begin{tabular}{ccccc}
	  		\hline
	  		\#Vocab & \#Doc & \#Query & \#Retrieval doc & Avg doc\\
	  		\hline
	  		0.6M & 0.5M & 250 & 2000 & 477\\
	  		\hline 	
		\end{tabular}
	\end{table}
	
	\begin{figure*}
		\centering
		\subfigure[Dot Product]{
			\label{Fig.sub.1}
			\includegraphics[width=0.31\textwidth]{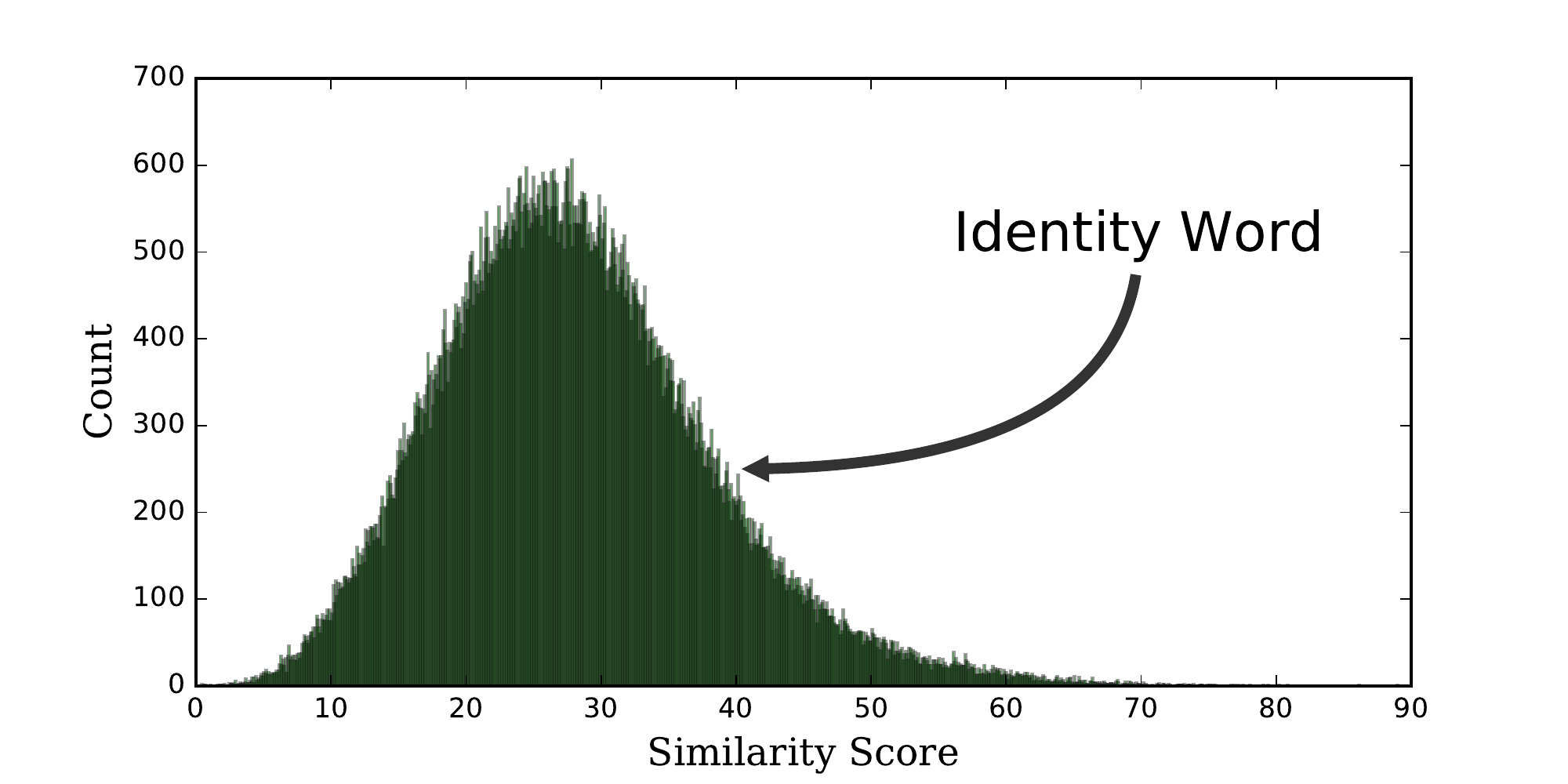}
			}
		\subfigure[Cosine]{
			\label{Fig.sub.3}
			\includegraphics[width=0.31\textwidth]{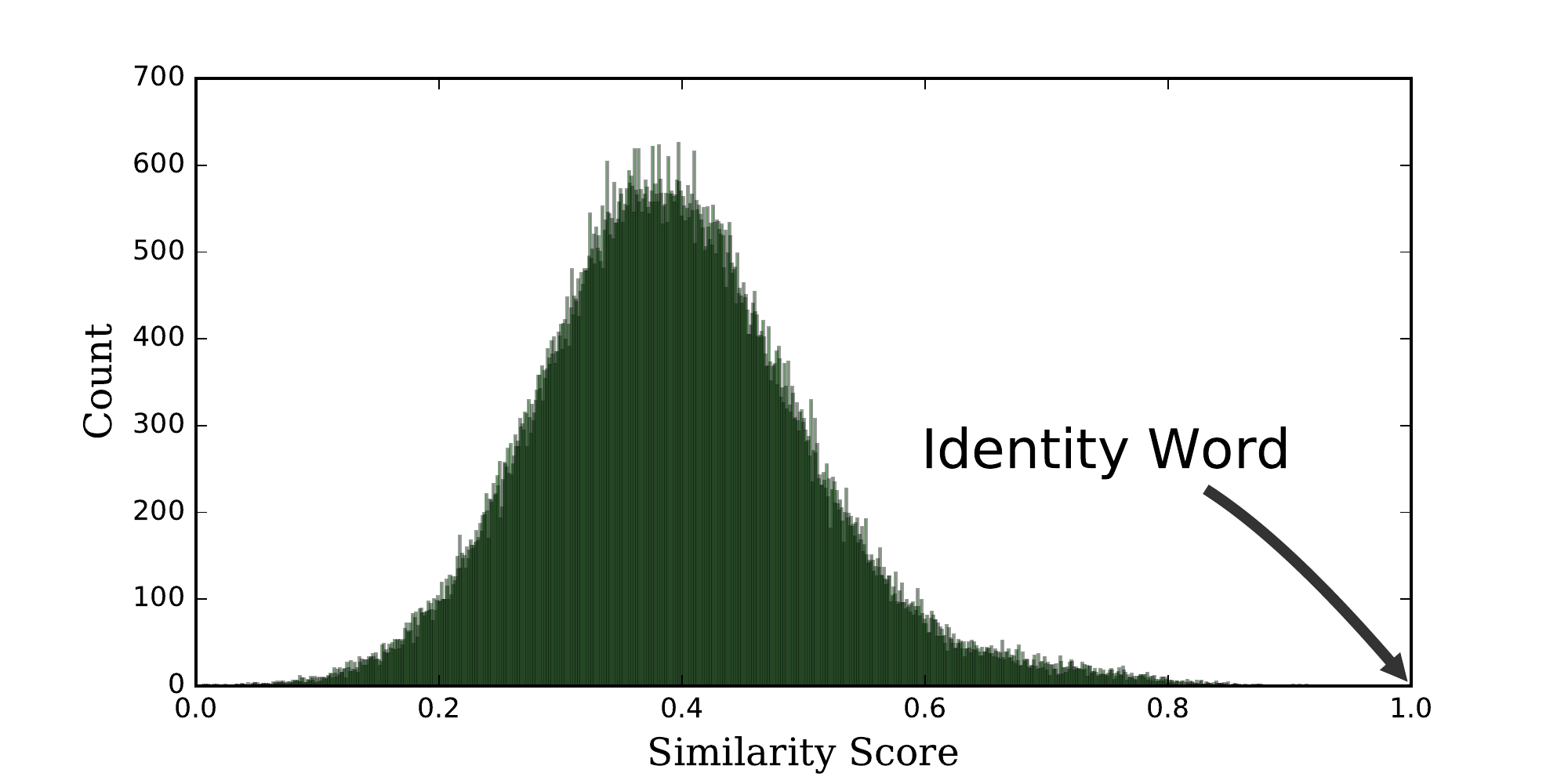}
 		}
 		\subfigure[Gaussian Kernel]{
			\label{Fig.sub.3}
			\includegraphics[width=0.31\textwidth]{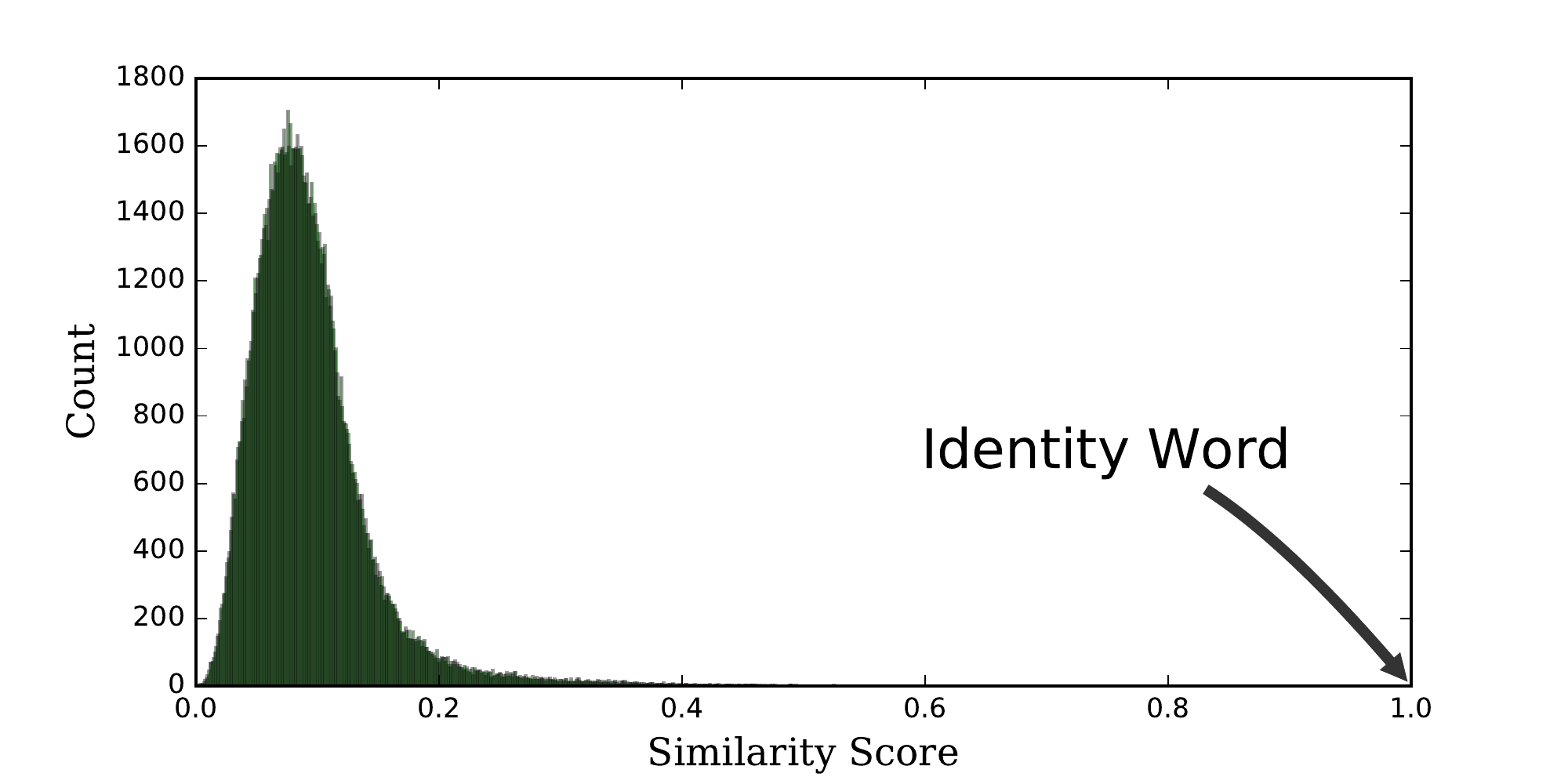}
 		}
 		\caption{Choose one word from the vocabulary and measure the similarity between other words, we draw the histogram of three type of similarity functions: dot product, cosine and gaussian kernel. The arrow point the similarity between two identity word (the word we choose).}
 		\label{Fig.word_dist}
	\end{figure*}
	
\subsection{Detailed Studies on MatchPyramid Models for Ad-hoc Retrieval}

	\subsubsection{Impact of Pooling Size}
	We first study the effect of pooling size. Pooling layers are used to reduce the dimension of the feature maps, and to pick out the most important information for the latter layers. Especially in ac-hoc retrieval task, documents often contain hundreds of words, but most of them might be background words. So the pooling layers might be even more important to distill the useful information from the noisy background. In this experiment, we try different pooling sizes and show the results in Table~\ref{Table.4}. In our experiments, the maximum query length is 5 and we truncate document length to 500, for the computational efficiency. Actually, we have tried other length of document, such as 1000 or full length, but the performance change is slight.
	
	\begin{table}[[!htbp]
		\centering
		\caption{Comparison of different pooling size of MatchPyramid. (fix kernel size $1 \times 1$)\label{Table.4}}
		\begin{tabular}{lllll}
	  		\hline
	  		Model & Pooling Size & MAP & nDCG@20 & P@20\\
	  		\hline	  		
	  		MP-Ind & $5 \times 100$ & 0.175 & 0.320 & 0.254 \\
	  		MP-Ind & $5 \times 50$ & 0.195 & 0.343 & 0.266 \\
	  		MP-Ind & $5 \times 20$ & 0.209 & 0.363 & 0.279 \\
	  		MP-Ind & $5 \times 10$ & \textbf{0.219} & \textbf{0.387} & \textbf{0.301} \\
	  		MP-Ind & $5 \times 5$ & 0.214 & 0.380 & 0.295 \\
	  		MP-Ind & $5 \times 3$ & 0.209 & 0.370 & 0.300 \\
	  		\hline
	  		MP-Ind & $3 \times 20$ & 0.213 & 0.357 & 0.285 \\
	  		MP-Ind & $3 \times 10$ & \textbf{0.225} & \textbf{0.387} & \textbf{0.302} \\
	  		MP-Ind & $3 \times 5$ & 0.225 & 0.385 & 0.301 \\
	  		MP-Ind & $3 \times 3$ & 0.215 & 0.377 & 0.301 \\
	  		\hline 	
	  		MP-Ind & $1 \times 10$ & 0.056 & 0.082 & 0.073 \\
	  		MP-Ind & $1 \times 5$ & 0.051 & 0.072 & 0.078 \\
	  		MP-Ind & $1 \times 3$ & 0.043 & 0.066 & 0.053 \\
	  		\hline
		\end{tabular}
	\end{table}
	
	Results show that too small or too large pooling size is not proper for this task. The best pooling size is about $3 \times 10$, which means on the query side, we use the median query length, and on the document side, we aggregate up every 50 words, close to the average length of a paragraph.

	\subsubsection{Impact of Similarity Function}
	Similarity function is used to measure the similarity of two words and build the matching matrix. For paraphrase identification task, the previous results show that indicator function is the best. For question answering, we find that dot product function is better than others. Here we try to explore the similarity function for ad-hoc retrieval task. We compare four kinds of similarity function: indicator function, dot product, cosine and gaussian kernel.
	
	\begin{table}[!htbp]
		\centering
		\caption{Comparison of different similarity function of MatchPyramid. (fix kernel size $1 \times 1$, pooling size $3 \times 10$)\label{Table.5}}
		\begin{tabular}{llll}
	  		\hline
	  		Model & MAP & nDCG@20 & P@20\\
	  		\hline	  		
	  		MP-Ind & 0.225 & 0.387 & 0.302 \\
	  		MP-Dot & 0.095 & 0.149 & 0.142 \\
	  		MP-Cos & 0.189 & 0.340 & 0.272 \\
	  		MP-Gau & \textbf{0.226} & \textbf{0.403} & \textbf{0.318} \\
	  		\hline 	
		\end{tabular}
	\end{table}
	
	As we can see, MP-Ind performs quite well, indicating that exact matching signals are very useful for the ad-hoc retrieval task. By using embeddings, we can see gaussian kernel similarity function is better than others. To understand the underlying reasons, we first take a look at words similarity distribution as shown in Figure~\ref{Fig.word_dist}. We can see that the exact matching score (arrow pointed) is the largest value under similarity function cosine and gaussian kernel, but not under dot product. So MP-Cos and MP-Gau keep the strength of exact matching signals, while MP-Dot loses this ability and leads to worse result. The performance gap between MP-Cos and MP-Gau may be related to the gap between exact matching and the semantic matching scores. The large gap in MP-Gau indicates that the model can better differentiate exact matching from semantic matching and this leads to better performance. 

\subsubsection{Impact of Kernel Size}
	In this section we study the effect of kernel size of the convolutional layer in the MatchPyramid model. We conduct the experiments on different sizes of kernel with MP-Ind and MP-Gau in this section, because MP-Dot and MP-Cos do not work well in ad-hoc retrieval task. The results are shown in Table~\ref{Table.3}.
	
	\begin{table}[!htbp]
		\centering
		\caption{Comparison of different kernel size of MatchPyramid. (fix pooling size $3 \times 10$)\label{Table.3}}
		\begin{tabular}{lllll}
	  		\hline
	  		Model & Kernel Size & MAP & nDCG@20 & P@20\\
	  		\hline	  		
	  		MP-Ind & $1 \times 1$ & 0.225 & \textbf{0.387} & \textbf{0.302} \\
	  		MP-Ind & $1 \times 3$ & \textbf{0.226} & 0.382 & 0.294 \\
	  		MP-Ind & $1 \times 5$ & 0.223 & 0.382 & 0.297 \\
	  		MP-Ind & $3 \times 3$ & 0.221 & 0.379 & 0.295 \\
	  		MP-Ind & $5 \times 5$ & 0.219 & 0.378 & 0.295 \\
	  		\hline
	  		MP-Gau & $1 \times 1$ & 0.226 & 0.403 & 0.318 \\
	  		MP-Gau & $1 \times 3$ & \textbf{0.232} & \textbf{0.411} & \textbf{0.327} \\
	  		MP-Gau & $1 \times 5$ & 0.226 & 0.409 & 0.326 \\
	  		MP-Gau & $3 \times 3$ & 0.220 & 0.400 & 0.312 \\
	  		MP-Gau & $5 \times 5$ & 0.201 & 0.371 & 0.301 \\
	  		\hline 	
		\end{tabular}
	\end{table}
	
	We have tried two kinds of kernel sizes, $1 \times n$ and $n \times n$, where $n \in [1, 3, 5]$. We expect kernel $1 \times n$ to capture the information around the central word in the kernel window. The results show that different kernel sizes under indicator similarity function perform similarly. It is reasonable if we look at the Matching Matrix generated by the indicator function. The Matching Matrix is very sparse and in a kernel window there is usually one non-zero element. So the kernel size is not that important for indicator function. However, by using the gaussian kernel, we introduce semantic word similarity into the model, and a proper kernel size will get more information and generate a better result. We find that MP-Gau with kernel size $1 \times 3$ achieves the best performance. Additionally, we expect kernel $n \times n$ to capture the word proximity information, such as n-gram matching. Surprisingly there is no improvement in these experiments. The reason might be that the dataset is too small to learn the complex proximity patterns.

\subsection{Comparison with Baseline Models}
	We further compare the MatchPyramid model with a set of baseline models. We adopt three types of baselines, including traditional models, representation-based deep matching models and interaction-based deep matching model.
	
	Traditional models include:
	
	\textbf{QL}: Query likelihood model based on Dirichlet smoothing\cite{zhai2001study} is one of the best language models.
	
	\textbf{BM25}: Based on BM25 formula \cite{robertson1994some}, is another highly effective retrieval model.
	
	Representation-based deep matching models include:
	
	\textbf{DSSM}: DSSM model \cite{DSSM} uses fully connected layers to encode query and document into two fix-length vectors, then uses cosine similarity to compute the matching score. Since DSSM needs large scale training data due to its huge parameter size, we directly used the released model\footnote{http://research.microsoft.com/en-us/downloads /731572aa-98e4-4c50-b99d-ae3f0c9562b9/} (trained on large click-through dataset) in our experiments.
	
	\textbf{CDSSM}: CDSSM model \cite{CDSSM0} is similar with DSSM, but use convolutional layers to encode query and document. For the same reason as DSSM, we also made use of the released model directly.
	
	\textbf{ARC-I}: ARC-I model \cite{ARC-II} uses convolutional layers to encode two texts, and uses fully connected layers to aggregate matching score.
	
	Interaction-based deep matching models include:
	
	\textbf{ARC-II}: ARC-II \cite{ARC-II} constructs local interactions by adding up word embeddings in a small context window, then makes use of convolutional layers to extract features from the interactions.

	\begin{table}[!htbp]
		\centering
		\caption{Comparison of different retrieval models over the TREC collection Robust04. $^{\dagger}$ models trained on large click-through dataset.\label{Table.2}}
		\begin{tabular}{lllll}
	  		\hline
	  		Type & Name & MAP & nDCG@20 & P@20\\
	  		\hline
	  		Traditional & QL & 0.253 & 0.415 & 0.369\\
	  		Model & BM25 & \textbf{0.255} & \textbf{0.418} & \textbf{0.370}\\
	  		\hline
	  		Representation & DSSM$^{\dagger}$ & 0.095 & 0.201 & 0.171 \\
	  		Based Model & CDSSM$^{\dagger}$ & 0.067 & 0.146 & 0.125 \\
	  		 & ARC-I & 0.041 & 0.066 & 0.065 \\
	  		\hline
	  		Interaction & ARC-II & 0.067 & 0.147 & 0.128 \\
	  		Based Model & MP-Gau & 0.232 & 0.411 & 0.327 \\
	  		\hline
		\end{tabular}
	\end{table}
	
	The experimental results (see Table~\ref{Table.2}) show that MatchPyramid outperforms all the representation-based deep matching models. The major reason is that it can retain all the low-level matching signals which are important for ad-hoc retrieval. For ARC-II, using 1D convolution to generate intercations signals in a small context window seems not very effective for matching. However, we find that the best performing deep matching model, MP-Gau, still cannot compete with traditional retrieval models. The results indicate that the ad-hoc retrieval task may be quire different from other text matching tasks, such as paraphrase identification and question answering. We need some further studies on the differences between these tasks for designing better deep matching models.

	
\section{Conclusions}
In this paper, we apply MatchPyramid model to ad-hoc retrieval task and discuss the impact of different kernel sizes, pooling sizes and similarity functions. We find that pooling by paragraph length in document, a good similarity function which can differentiate exact matching signals from semantic matching signals, and a relative small kernel size are helpful for the ad-hoc retrieval task. Experiments show that the MatchPyramid models can significantly outperform several recently introduced deep matching models on the retrieval task, but still cannot compete with the traditional retrieval models. These results encourage us to seek deeper understanding of the text matching task in ad-hoc retrieval and propose better models accordingly.


\bibliographystyle{abbrv}
\bibliography{sigir_work}  

\begin{thebibliography}{10}

\bibitem{giles2001overfitting}
R.~C. S. L.~L. Giles.
\newblock Overfitting in neural nets: Backpropagation, conjugate gradient, and
  early stopping.
\newblock In {\em Advances in Neural Information Processing Systems 13:
  Proceedings of the 2000 Conference}, volume~13, page 402. MIT Press, 2001.

\bibitem{ARC-II}
B.~Hu, Z.~Lu, H.~Li, and Q.~Chen.
\newblock Convolutional neural network architectures for matching natural
  language sentences.
\newblock In {\em Advances in Neural Information Processing Systems}, pages
  2042--2050, 2014.

\bibitem{DSSM}
P.-S. Huang, X.~He, J.~Gao, L.~Deng, A.~Acero, and L.~Heck.
\newblock Learning deep structured semantic models for web search using
  clickthrough data.
\newblock In {\em Proceedings of the 22nd ACM international conference on
  Conference on information \& knowledge management}, pages 2333--2338. ACM,
  2013.

\bibitem{kingma2014adam}
D.~Kingma and J.~Ba.
\newblock Adam: A method for stochastic optimization.
\newblock {\em arXiv preprint arXiv:1412.6980}, 2014.

\bibitem{DeepMatch}
Z.~Lu and H.~Li.
\newblock A deep architecture for matching short texts.
\newblock In {\em Advances in Neural Information Processing Systems}, pages
  1367--1375, 2013.

\bibitem{MatchPyramid}
L.~Pang, Y.~Lan, J.~Guo, J.~Xu, S.~Wan, and X.~Cheng.
\newblock Text matching as image recognition.
\newblock {\em CoRR}, abs/1602.06359, 2016.

\bibitem{CNTN}
X.~Qiu and X.~Huang.
\newblock Convolutional neural tensor network architecture for community-based
  question answering.
\newblock In {\em Proceedings of the 24th International Joint Conference on
  Artificial Intelligence (IJCAI)}, pages 1305--1311, 2015.

\bibitem{robertson1994some}
S.~E. Robertson and S.~Walker.
\newblock Some simple effective approximations to the 2-poisson model for
  probabilistic weighted retrieval.
\newblock In {\em Proceedings of the 17th annual international ACM SIGIR
  conference on Research and development in information retrieval}, pages
  232--241. Springer-Verlag New York, Inc., 1994.

\bibitem{CDSSM0}
Y.~Shen, X.~He, J.~Gao, L.~Deng, and G.~Mesnil.
\newblock A latent semantic model with convolutional-pooling structure for
  information retrieval.
\newblock In {\em Proceedings of the 23rd ACM International Conference on
  Conference on Information and Knowledge Management}, pages 101--110. ACM,
  2014.

\bibitem{uRAE}
R.~Socher, E.~H. Huang, J.~Pennin, C.~D. Manning, and A.~Y. Ng.
\newblock Dynamic pooling and unfolding recursive autoencoders for paraphrase
  detection.
\newblock In {\em Advances in Neural Information Processing Systems}, pages
  801--809, 2011.

\bibitem{zhai2001study}
C.~Zhai and J.~Lafferty.
\newblock A study of smoothing methods for language models applied to ad hoc
  information retrieval.
\newblock In {\em Proceedings of the 24th annual international ACM SIGIR
  conference on Research and development in information retrieval}, pages
  334--342. ACM, 2001.

\end{thebibliography}
%
%
\end{document}